# QUALITY OF SERVICE PROVISIONING IN MANET USING A CROSS-LAYER APPROACH FOR ROUTING

Ruchita Goyal, Divyanshu and Manoj Mishra

Department of Electronics and Computer Engineering,
Indian Institute of Technology Roorkee, India
divyanshu26@gmail.com

## ABSTRACT

*Deployment of multimedia applications warrants provisioning of Quality of Service (QoS) in MANET. However, limited battery power, other resource constraints and mobility of nodes make QoS provisioning difficult to achieve in MANET. This difficulty can be overcome by using a cross-layer approach for routing. In [1] Patil et al., proposed a cross-layer routing protocol named Cost Based Power Aware Cross Layer – AODV (CPACL-AODV) which overcomes the limitation of battery power of nodes. Though many similar energy efficient and cross-layer routing protocols have been proposed for MANET, none of them handles QoS. A novel MANET routing protocol, Type of Service, Power and Bandwidth Aware AODV (TSPBA-AODV), which overcomes resource constraints and simultaneously provides QoS guarantees using a cross-layer approach, is proposed in this paper. In addition the effect of variation in nodes' mobility on performance of TSPBA-AODV is compared with that of CPACL-AODV [1] for two different types of network traffic. As shown by the results of simulations performed, TSPBA-AODV performs better than CPACL-AODV for MANET in which nodes move with small speeds (speeds up to 40 Km/hr approx.). In addition the effect of variation in data sending rate of nodes on performance of the protocols is also studied. As shown by the results of simulations performed, TSPBA-AODV performs better than CPACL-AODV for all variations in data sending rate of nodes.*

## 1. INTRODUCTION

Highly dynamic nature and varied usefulness of MANET have provoked great interest in research community for its study. MANET has a unique capability of forming an infrastructure-less network on the fly without need of any central administration. Earlier the use of MANET was limited to emergency situations such as natural disasters, military conflicts, emergency medical facilities, etc. However, proliferation of multimedia applications over internet has also affected MANET and warranted provisioning of QoS in it. This requirement can be met by designing a cross-layer routing protocol in such a way that QoS can be provisioned. Earlier the cost function used for making routing decisions in MANET was hop count. However since, MANET works under various constraints, such as limited battery power and frequently changing topology, the routing protocol can be effective only if it is designed by taking care of cross-layer parameters such as battery lifetime, application type, bandwidth, etc. Several cross-layer aware routing cost functions have been proposed by researchers for routing in MANET. Such cost functions can be divided into three types:

(1) Type 1 - *Monotonically increasing function*: A cost function which increases monotonously with time for a node. For example fraction of initial battery power utilized, etc.

(2) Type 2 – *Fluctuating function*: A cost function which doesn't show monotonic variation, but rather shows fluctuations by either increasing or decreasing. For example bandwidth utilization, fraction of message queue occupied, etc.

(3) Type 3 – *Hybrid function*: A cost function which is combination of Type 1 and Type 2 cost functions.

If nodes are taken to be identical and overall cost function of path is calculated as sum of cost function of nodes along a path (taking average is not much justified as a very long path can have better average), then irrespective of cost function chosen hop count will implicitly decide routes in the beginning because cost function for individual nodes will be almost in same range initially. However in case of cost function of the Type 1, due to monotonous increase in cost function with time, at a later stage routing will no more be dependent on hop count, but this can't assure quality of service. For example - a path with better sum of battery utilization can be longer and congested compared to a path with sustainable battery power but having lesser congestion or length and so lesser delay. On the other hand if cost function is taken to be of the second type then we are actually ignoring the resource constraints of MANET such as limited power supply.

In this paper we propose a novel protocol named Type of Service, Power and Bandwidth Aware AODV (TSPBA-AODV) that uses a Type 3 cost function. The scenario we are assuming is that we have a large No. of mobile nodes in a small area, so that hop count is not a matter of concern but rather QoS provisioning in dynamic environment is the matter of concern. Earlier most of the protocols used either Type 1 or Type 2 cost function. Recently a protocol named CPACL-AODV [1] was proposed by Patil et al., which uses cost function of Type 1, based on battery utilization. We compared performance of our protocol (TSPBA-AODV) with CPACL-AODV (implemented with assumption that energy loss due to transmission or receipt is directly proportional to size of the packet dealt with).

Rest of the paper is organized as follows. Section 2 gives a brief overview of related works. Section 3 describes TSPBA-AODV. Section 4 discusses about the simulation environment used, performance metrics evaluated and the results obtained. Finally, section 5 concludes our paper.

## 2. RELATED WORKS

Extensive research has been done in finding the most suitable cost function for routing in MANET. Table 1 classifies recent cross-layer routing protocols on the basis of type of cost function (types as described in section 1) used.

Table 1.  Classification of recent cross-layer routing protocols on the basis of type of cost function used.

| Type 1 | Type 2 | Type 3 |
|---|---|---|
| SNR/RP aware routing protocol [3] in case RP is used | SNR/RP aware routing protocol [3] in case SNR is used | TSPBA-AODV (discussed in section 3) |
| CPACL–AODV [1] | ALW protocol [6] | TEA – AODV [5] |
| Protocol proposed in [2] | TSLA protocol [7] | |
| SQ–AODV [4] | | |

Earlier a cost function that accounted for the utilization of battery of a node was proposed by Patil et al., [1]. However Patil et al., assumed that transmit power of a node is fixed which doesn't occur in real life situation. Power lost due to transmission depends upon the size of the message transmitted. Moreover they accounted only for battery power during routing, but it is possible that a route that has better battery life can have high delay due to high congestion or more length, whereas a route with sustainable battery power can provide better delay. Moreover the path selected by considering battery life only may not be capable of providing minimum bandwidth guarantee to multimedia applications. So this protocol is not suitable for multimedia applications. In another protocol, the authors Enneya et al., accounted for mobility only in their cost function for routing [2]. However, it is possible that a bit more mobile path than the least mobile path can give better bandwidth and delay guarantees. Moreover they didn't consider the major constraint of limitation of battery power of nodes. Alnajjar et al., proposed a protocol that uses either SNR or remaining power (RP) to determine route [3]. Accounting only for SNR can't provide bandwidth and delay guarantees required by multimedia applications. Moreover it accounts either for SNR or RP but not for both. Veerayya et al., proposed a protocol that accounts for battery power available at nodes for routing [4]. Thus, only Type 1 cost function is used, ignoring the QoS requirements of applications. Pushpalatha et al., proposed a protocol which uses trust based on forwarding frequency as a cost function [5]. However, trust alone can't guarantee QoS provisioning. Al-Khwildi et al., proposed a protocol that considers cost function on the basis of type of application [6]. However, the protocol is entirely new rather than being an extension to an existing protocol. Moreover as per Al-Khwildi et al., the protocol provides advantage over AODV only in terms of network load and route discovery time. Mbarushimana et al., proposed a protocol that takes into account the type of service and load in network while routing [7], both these parameters give the cost function corresponding to Type 2 and ignore the limitation of battery power of nodes.

## 3. TYPE OF SERVICE, POWER AND BANDWIDTH AWARE AODV (TSPBA-AODV)

TSPBA-AODV is an extension to AODV. In this section the changes made over AODV to design TSPBA-AODV are discussed.

TSPBA-AODV divides applications into following two categories on the basis of their requirements:

(1) Type-1 – Loss tolerant applications that are not delay sensitive. For example – DNS service, etc.

(2) Type-2 – Loss tolerant applications that are delay sensitive. For example – instant messaging, etc.

Each node maintains two different routing tables Table-1 and Table-2 for Type-1 and Type-2 types of applications, respectively. For an application, the type of application is decided from the above two types and then corresponding routing table and corresponding control packets are used. However, algorithm used is same for both types of application.

### 3.1. Format of Packets and Route Table Entry

New fields are added to RREQ, RREP, RERR and route table entry. They are as follows:

1) For RREQ packet:

   a) Boolean field named isComputeCost, whose purpose is described in subsection 3.2

   b) Integer field named appType to identify the type of application (Type-1 or Type-2) to which RREQ packet belongs.

   c) Double field named cumulativeCost, which is updated from hop to hop. It denotes the cost of route from the origin to the hop forwarding the RREQ.

2) For RREP packet:

a) Integer field named appType to identify the type of application (Type-1 or Type-2) to which RREP packet belongs.

b) Double field named cumulativeCost, which is kept constant from hop to hop. It denotes the cost of route from origin to destination.

3) For RERR packet:

a) Integer field named appType to identify the type of application (Type-1 or Type-2) to which RERR packet belongs.

4) For route table entry:

a) Double field named cumulativeCost, which represents cost of the route.

Here wherever cost of a route is talked about it is the sum of costs of each node in the route including end points of it also.

### 3.2. Route Discovery and Maintenance

The protocol is reactive. Whenever a node has to send data to another node it searches the route table corresponding to the application for entry corresponding to the destination. If the entry is found it sends the packet to the corresponding next hop. If entry is not found then it broadcasts RREQ packet with isComputeCost field reset, cumulativeCost set to cost of the source node corresponding to the type of application and appType given the value corresponding to the application type. The algorithm followed by a node receiving a RREQ packet can be divided into four steps: request filtration step, source entry renewal step, reply generation step and request propagation step. These steps are discussed in the following subsections by means of flowcharts.

#### 3.2.1. Request Filtration Step

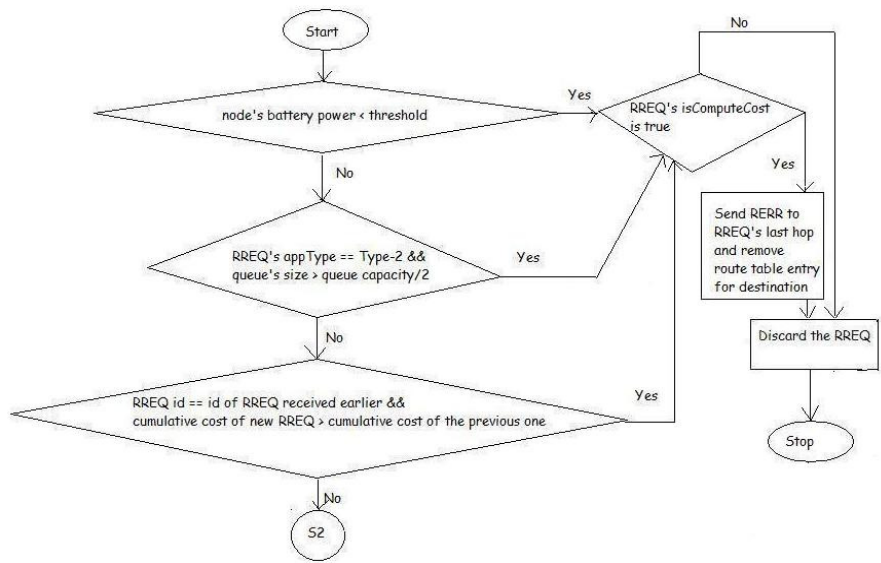

### 3.2.2. Source Entry Renewal Step

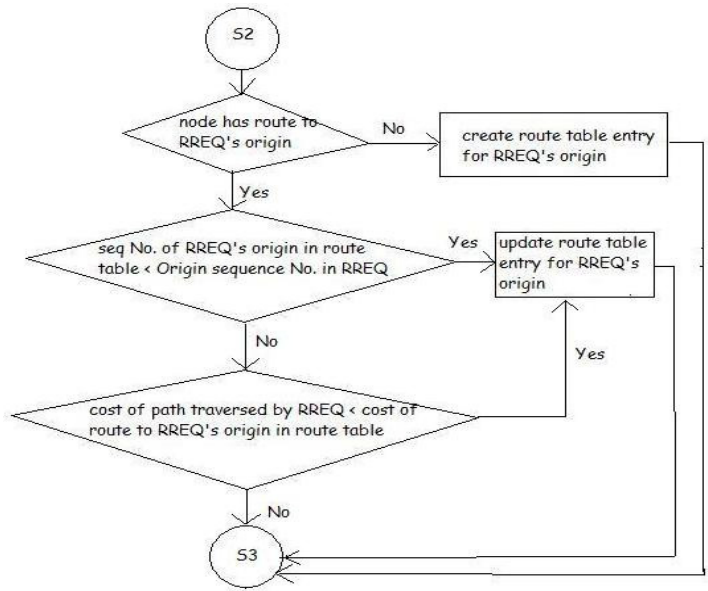

### 3.2.3. Reply Generation Step

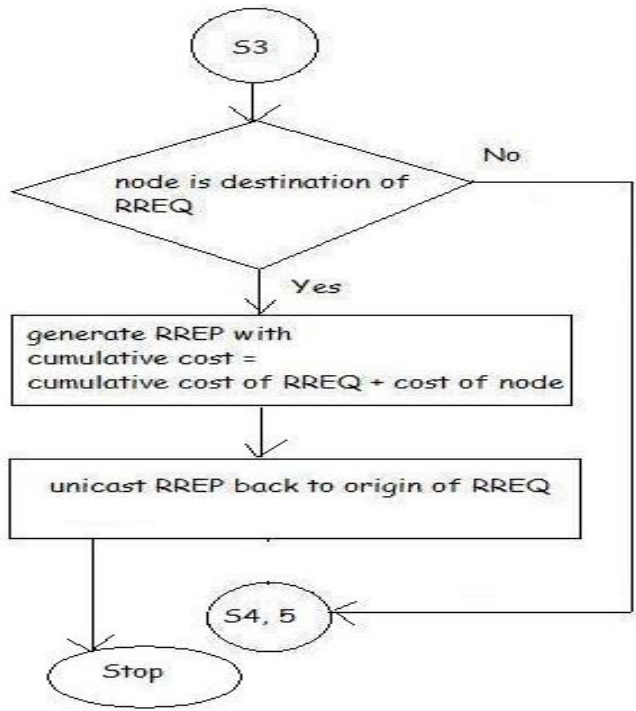

**3.2.4. Request Propagation Step**

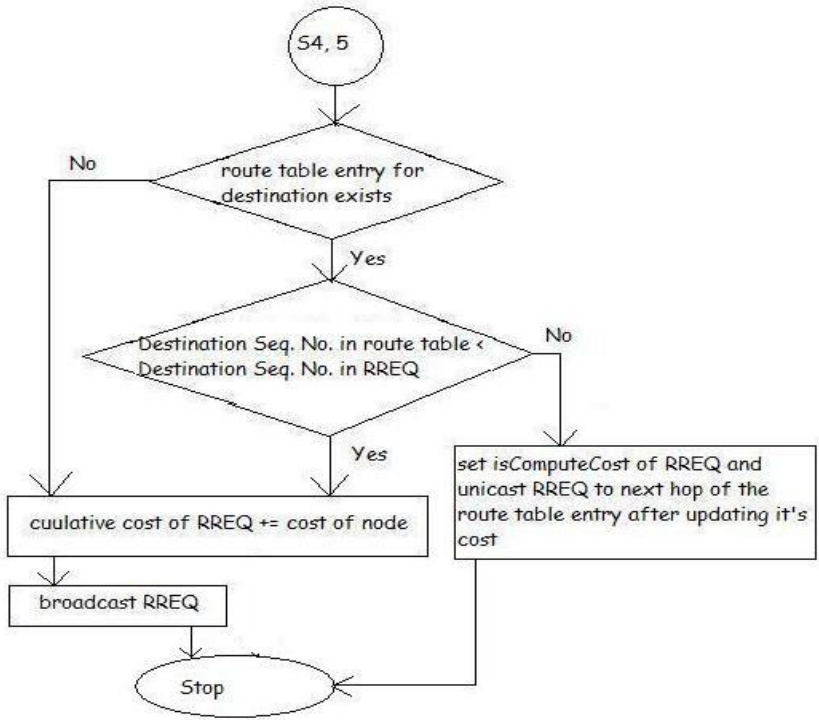

## 3.3. Cost Calculation

Cost of a node has following three components:
1) BandwidthCost ($B_i(t)$) – Utilization of radio of node-i at time t.
2) DelayCost ($D_i(t)$) – Fraction of queue of node-i occupied at time t.
3) PowerCost ($P_i(t)$) – Fraction of initial power that has been consumed till time t at node-i.
Cost of a node is weighted mean of all three of these costs but the weightings depend upon the type of application in the way shown in Table 2. The values of the weightings were decided experimentally in order to optimize performance.

Table 2. Weightings related to components depending upon the type of application.

| Application type / Weighting for component | b | d | p |
|---|---|---|---|
| Type-1 | 0.33 | 0.33 | 0.33 |
| Type-2 | 0.30 | 0.40 | 0.30 |

Overall cost of a node at time t, i.e., $C_i(t)$ , is given by (1) as

$$C_i(t) = b* B_i(t) + d* D_i(t) + p* P_i(t). \tag{1}$$

$$C(\pi, t) = \sum_{i \in \pi} C_i(t) \tag{2}$$

$C(\pi, t)$ in (2) represents cost of a route and is likely to be least for the path having minimum hop count because each of the cost components is a proper fraction. The protocol assumes that the field has a lot of nodes and the basic requirement is to effectively provision QoS. This assumption is made in order to ensure that change in one cost component may not subvert the change in other cost components. Thus, there are multiple routes between a source and a destination, each with almost similar hop count (around the minimum hop count), but with different QoS parameters and the problem is to find the most efficient path which can provision QoS.

If there are 2 routes with costs $C_1(\pi, t)$ and $C_2(\pi, t)$ such that $C_j(\pi, t) = b*\sum B_{ij}(t) + d*\sum D_{ij}(t) + p*\sum P_{ij}(t)$ for j=1, 2, then comparison between $C_1(\pi, t)$ and $C_2(\pi, t)$ not only signifies comparison between the overall cost but it also signifies comparison between $\sum B_{i1}(t)$ and $\sum B_{i2}(t)$, i.e., there is emphasis on $\triangle(\sum B_i(t))$ and similarly on $\triangle$ values of other cost components. So, for delay sensitive traffic if more weighting is given to $D_i(t)$ then it implicitly means that more weighting is given to $\triangle(\sum D_i(t))$, so the emphasis is more on difference created by $\triangle(\sum D_i(t))$ than on that created by any other cost component. Thus, priority is granted to our most important cost component. To account for the case in which path corresponding to $C_1(\pi, t)$ has less battery power left than the path corresponding to $C_2(\pi, t)$, PowerCost component of $C_1(\pi, t)$ will become high and $C_1(\pi, t)$ will not be comparable to $C_2(\pi, t)$, until its other components give better cost. Similarly, the case in which path corresponding to $C_1(\pi, t)$ has less available bandwidth than the path corresponding to $C_2(\pi, t)$ is taken care of by BandwidthCost.

## 4. SIMULATIONS AND RESULTS

All simulations were performed using JiST-SWANS-1.0.6. For each scenario 10 simulations were run and the result for the scenario was taken as average of the results of these 10 runs.

### 4.1. Simulation Environment

Sixty nodes were taken in a 2D field of 1000m *1000m. Each of the simulations was run for 600 sec. The initial battery capacity of each node was taken as 100 units. This initial energy was progressively reduced by data transmission/reception. When it reached zero units for a node, then the node could no more take part in the communication and was regarded as dead. The mobility model used was random waypoint with pause time as 10 sec and granularity as 10 sec. Rest of the parameters were varied to assess performances of protocols as mentioned in subsection 4.3.

### 4.2. Performance Metrics

Four performance metrics were used to compare the protocols. They are as follows:

1) Average end to end delay - It includes all delays possible namely, transmission delay, processing delay, queuing delay and propagation delay.

2) Throughput - This metric gives the number of bits that are successfully delivered to corresponding destinations in unit time in the network.

3) Packet Delivery Ratio - It is the ratio of the number of data packets successfully delivered to the destinations to the number of data packets generated by the sources.

4) Control Overhead - The number of routing packets transmitted per data packet delivered at the destination. Each hop wise transmission of a routing packet is counted as one transmission.

## 4.3. Results

This section discusses and compares the performances of the reference protocol (CPACL-AODV) and the proposed protocol (TSPBA-AODV).

### 4.3.1. Effect of Variation in Node Speed

Rate of sending data packets was kept at 1 packet/sec/node. Range of speeds of nodes was varied and performances of the two protocols were compared for two different types of traffic. This section deals with these comparisons.

**UDP Traffic**

The traffic used for simulation in this subsection was taken to be loss tolerant and delay insensitive with packets arriving in the network at constant bit rate.

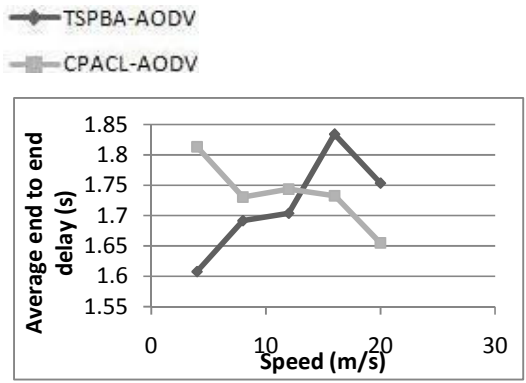

Figure 1.  Delay vs. Speed for UDP Traffic

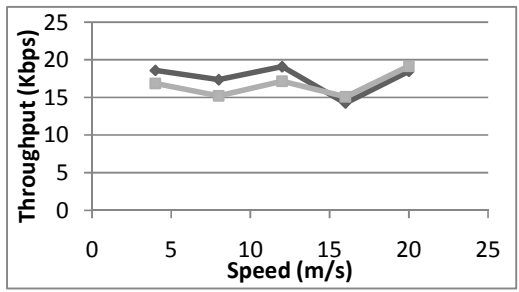

Figure 2.  Throughput vs. Speed for UDP Traffic

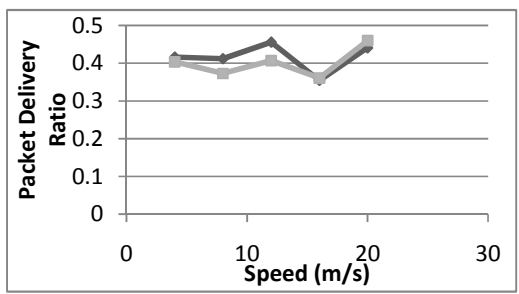

Figure 3.  Packet Delivery Ratio vs. Speed for UDP Traffic

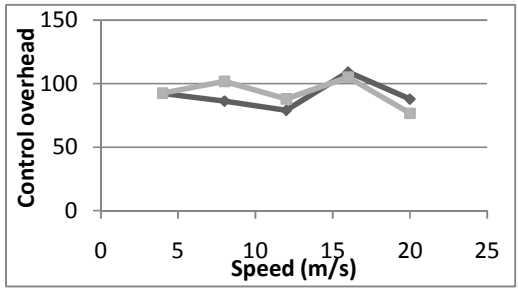

Figure 4.  Control Overhead vs. Speed for UDP Traffic

As it can be seen in Fig. 1 to Fig. 4, maximum percentage improvement of TSPBA-AODV over CPACL-AODV are 11.32% for delay, 14.02% for throughput, 12.04% for packet delivery ratio, and 15.25% for control overhead.

**UDP Delay Sensitive Traffic**
The traffic used for simulation in this subsection was taken to be loss tolerant and delay sensitive with packets arriving in the network at constant bit rate.

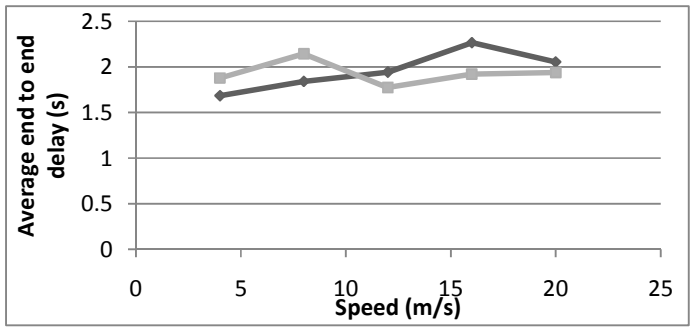

Figure 5.  Delay vs. Speed for UDP Delay Sensitive Traffic

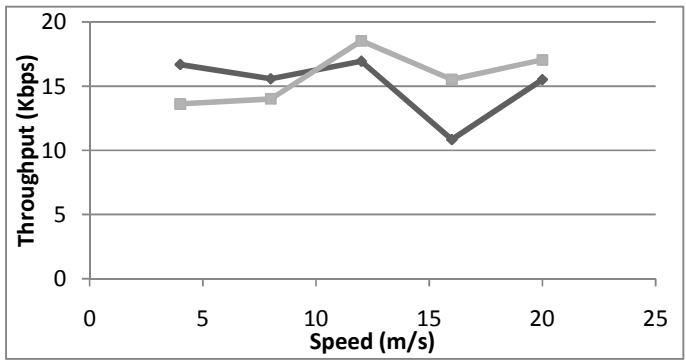

Figure 6.  Throughput vs. Speed for UDP Delay Sensitive Traffic

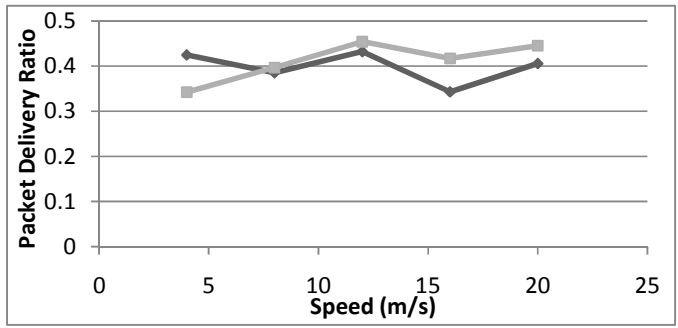

Figure 7. Packet Delivery Ratio vs. Speed for UDP Delay Sensitive Traffic

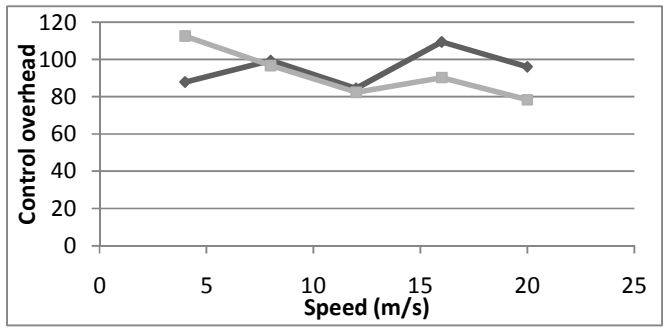

Figure 8. Control Overhead vs. Speed for UDP Delay Sensitive Traffic

As it can be seen in Fig. 5 to Fig. 8, maximum percentage improvement of TSPBA-AODV over CPACL-AODV are 14.08% for delay, 22.55% for throughput, 24.07% for packet delivery ratio, and 21.92% for control overhead.

As it can be seen in Fig. 1 to Fig. 8, at low speeds performance of TSPBA-AODV is better than that of CPACL-AODV for UDP traffic, UDP delay sensitive traffic and UDP burst traffic because the cost function used for selecting the optimal path in case of TSPBA-AODV comprises of delay, bandwidth and power cost components, due to which the route chosen has following features:

(1) It has optimal value of power cost. So the route lasts long. Since the route lasts long, number of control packets is less in the network and so the control overhead is less. Long life of selected routes also decreases average end to end delay because a route once established lasts long and less time is wasted in maintenance and rediscovery of route between same source and destination nodes.

(2) It has optimal value of delay cost. So the path chosen has optimal congestion along it, due to which there is decrease in average end to end delay. Moreover there will be less packet loss due to congestion and so packet delivery ratio and throughput will increase.

(3) It has optimal value of bandwidth cost. So path chosen has optimal potential to deliver data, due to which there is increase in throughput, packet delivery ratio and decrease in average end to end delay.

However, the performance of the TSPBA-AODV degrades when nodes are highly mobile (speed more than 40 Km/hr approx.). This happens because when nodes are highly mobile then route breakage occurs frequently and the nodes along the established routes send control packets for route maintenance frequently, resulting in less availability of bandwidth, message queue and power for nodes along established routes which thereby results in selection of otherwise non-optimal route.

### 4.3.2. Effect of Variation in Data Sending Rate of Nodes

In another set of simulations speeds of nodes were kept between 2m/s and 6m/s, the rate of sending data packets was varied and performances of the two protocols were compared for UDP traffic. This section deals with these comparisons.

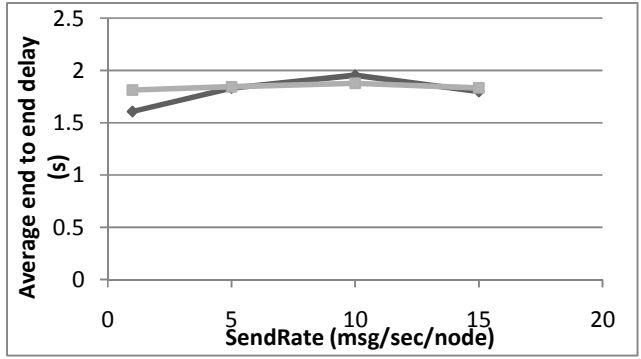

Figure 9.  Delay vs. SendRate for UDP Traffic

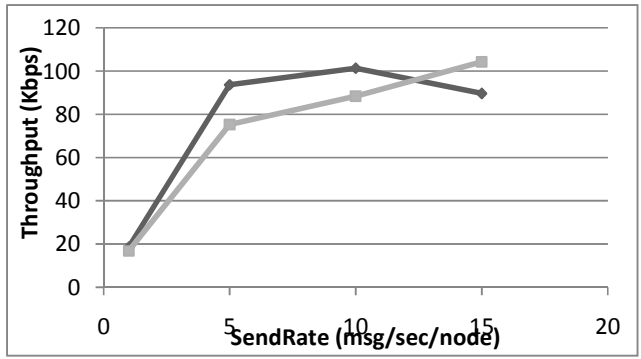

Figure 10.  Throughput vs. SendRate for UDP Traffic

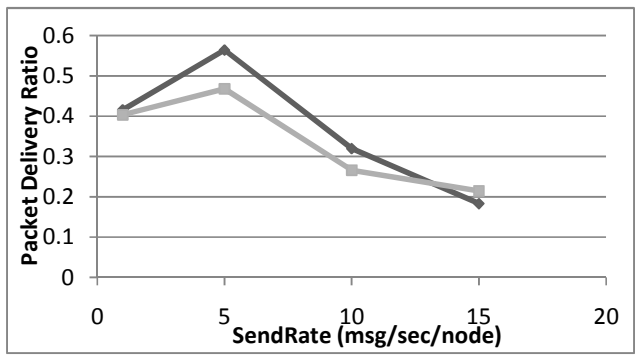

Figure 11.  Packet Delivery Ratio vs. SendRate for UDP Traffic

As it can be seen in Fig 9 to Fig 12, maximum percentage improvement of TSPBA-AODV over CPACL-AODV are 11.32% for delay, 24.48% for throughput, 20.55% for packet delivery ratio, 10.95% for control overhead. The reasons for this performance difference are same as the ones discussed in section 4.3.1.

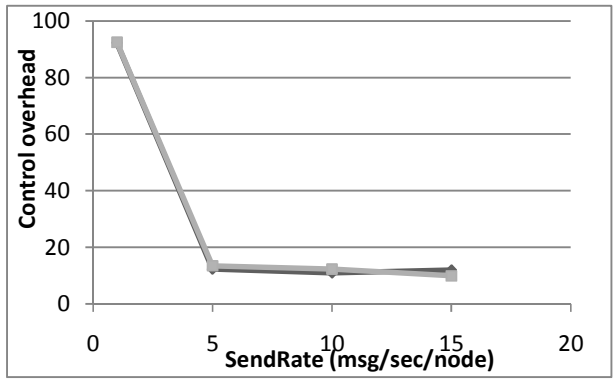

Figure 12. Control Overhead vs. SendRate for UDP Traffic

## 5. CONCLUSION

In this paper we proposed TSPBA-AODV, a cross-layer MANET routing protocol capable of provisioning QoS in MANET. Extensive simulations were performed. The results show that TSPBA-AODV performs better than CPACL-AODV [1] in case the speeds of nodes in the MANET are small (up to approx. 40 Km/hr). The performance is improved in terms of average end to end delay, throughput, packet delivery ratio and control overhead for two types of traffic: UDP traffic and UDP delay sensitive traffic. However, the performance of the proposed protocol degrades in scenario in which nodes are highly mobile (speed more than 40 Km/hr). So, the proposed protocol is better than the reference protocol for applications where mobility of nodes is small (up to 40 Km/hr). In addition the results of simulations show that TSPBA-AODV performs better than CPACL-AODV for all variations in data sending rate of nodes.

## REFERENCES

[1]  Patil, R. & Damodaram, A. (2008) "Cost Based Power Aware Cross Layer Routing Protocol For Manet", *IJCSNS International Journal of Computer Science and Network Security*, Vol. 8, No. 12, pp388-393.

[2]  Enneya, N., Koutbi, M.E. & Berqia, A. (2006) "Enhancing AODV Performance based on Statistical Mobility Quantification", *IEEE xplore*, pp2455-2460.

[3]  Alnajjar, F. & Chen, Y. (2009) "SNR/RP Aware Routing Algorithm: Cross-Layer Design for MANETS", *International Journal of Wireless & Mobile Networks (IJWMN)*, Vol. 1, No. 2, pp127-136.

[4]  Veerayya, M., Sharma & V., Karandikar, A. (2008) "SQ-AODV: A Novel Energy-Aware Stability-based Routing Protocol for Enhanced QoS in Wireless Ad-hoc Networks", *IEEE*, pp1-7.

[5]  Pushpalatha, M., Venkataraman, R. & Ramarao, T. (2009) "Trust Based Energy Aware Reliable Reactive Protocol in Mobile Ad Hoc Networks", *World Academy of Science, Engineering and Technolog*, 56.

[6]  Al-Khwildi, A.N., Khan, S., Loo, K.K., Al-Raweshidy, H.S.(2007) "Adaptive Link-Weight Routing Protocol using Cross-Layer Communication for MANET", *WSEAS Transactions on Communications*, Vol. 6, Issue. 11, pp833-839.

[7]  Mbarushimana, C. & Shahrabi, A.(2008) "TSLA: A QoS-Aware On-Demand Routing Protocol for Mobile Ad Hoc Networks", *Springer-Verlag Berlin Heidelber, LNCS 519*, pp265-278.